\documentclass[runningheads]{svmult}

\usepackage{makeidx}   
\usepackage{graphicx}  
\usepackage{subeqnar}  
\usepackage{multicol}  
\usepackage{physprbb}  
\makeindex             

\def\pichJournal#1#2#3#4{#1 {\bf #2}, #4 (#3)}
\begin{document}
\title*{The Standard Model of Particle Physics:\protect\newline 
Status \& Low-Energy Tests}
\toctitle{The Standard Model of Particle Physics:
\protect\newline Status \& Low-Energy Tests}
%
%
\titlerunning{The Standard Model of Particle Physics: 
Status \& Low-Energy Tests}
%
\author{Antonio Pich}
\authorrunning{Antonio Pich}
%
%
\institute{IFIC, Universitat de Val\`encia -- CSIC,
Apt. 22085, E-46071 Val\`encia, Spain}

\maketitle              

\begin{abstract}\index{abstract}
Precision measurements of low-energy observables provide 
stringent tests of the Standard Model structure and
accurate determinations of its parameters.
An overview of the present experimental status is presented.
The main topics discussed are the muon anomalous magnetic moment,
the asymptotic freedom of strong interactions, the lepton universality
of gauge couplings, the quark flavour structure and CP violation.
\end{abstract}

\vspace*{12cm}\hspace*{-.8cm}
\raisebox{-6.95em}[0pt][0pt]{\bf\small Invited talk at the
ESO--CERN--ESA Symposium on Astronomy, Cosmology}
\raisebox{-7em}[0pt][0pt]{\bf\small and Fundamental
Physics (Garching bei M\"unchen, Germany, March 2002)}
\vspace*{-13.85cm}

\section{Standard Model Structure}

The Standard Model (SM) \cite{EWSM,QCD}
is a gauge theory, based on the group
$SU(3)_C \otimes SU(2)_L \otimes U(1)_Y$,
which describes strong, weak and electromagnetic interactions,
via the exchange of the corresponding spin-1 gauge fields:
8 massless gluons and 1 massless photon for the strong and
electromagnetic forces, respectively,
and 3 massive bosons, $W^\pm$ and $Z$, for the weak interaction.
The gauge symmetry determines the dynamics in terms of the three
couplings $g_s$, $g$ and $g'$, associated with the $SU(3)_C$,
$SU(2)_L$ and $U(1)_Y$ subgroups. 
Strong interactions are governed by the first group factor,
while the other two provide a unified description of the
electroweak forces, their gauge parameters being related through
$g\, \sin{\theta_W} = g'\, \cos{\theta_W} = e$.

The fermionic matter content is given by the known
leptons and quarks, which are organized in a 3-fold
family structure:
\begin{equation}\label{eq:families}
\left[\begin{array}{cc}\nu_e & u \\ e^- & d'\end{array}\right]\; ,
\qquad\quad
\left[\begin{array}{cc}\nu_\mu & c \\  \mu^- & s'\end{array}\right]\; ,
\qquad\quad
\left[\begin{array}{cc}\nu_\tau & t \\  \tau^- & b'\end{array}\right]\; , 
\end{equation}
where
(each quark appears in 3 different {\it colours}) 
\begin{equation}\label{eq:structure}
\left[\begin{array}{cc}\nu_l & q_u \\ l^- & q_d \end{array}\right]\;\equiv\;
\left(\begin{array}{c} \nu_l \\ l^-\end{array}\right)_{\! L}\; ,\quad
\left(\begin{array}{c} q_u \\ q_d\end{array}\right)_{\! L}\; ,\quad l^-_R\; , 
\quad (q_u)_R\; , \quad
(q_d)_R\; ,
\end{equation}
plus the corresponding antiparticles.
Thus, the left-handed fields are $SU(2)_L$ doublets, while
their right-handed partners transform as $SU(2)_L$ singlets.
The three fermionic families in (\ref{eq:families}) appear
to have identical properties (gauge interactions); they only
differ by their mass and their flavour quantum number.

The gauge symmetry is broken by the vacuum,
which triggers the Spontaneous Symmetry Breaking (SSB)
of the electroweak group to the electromagnetic subgroup:
\begin{equation}\label{eq:ssb}
SU(3)_C \otimes SU(2)_L \otimes U(1)_Y \;\;
\stackrel{\mathrm{\scriptsize SSB}}{\longrightarrow}\;\;
SU(3)_C \otimes U(1)_{QED} \; .
\end{equation}
The SSB mechanism generates the masses of the weak gauge bosons,
and gives rise to the appearance of a physical scalar particle, 
the so-called {\it Higgs}. The fermion masses and mixings are 
also generated through the SSB mechanism.  

The SM constitutes one of the most successful achievements 
in modern physics. It provides a very elegant theoretical 
framework, which is able to describe all known experimental 
facts in particle physics. A detailed description of the SM 
and its impressive phenomenological success 
can be found in \cite{jaca:94,capri:97}.

\section{Quantum Corrections}

The high accuracy achieved by the most recent experiments 
allows to make stringent tests of the SM structure at the level 
of quantum corrections. The following discussion concentrates on
Quantum Electrodynamics (QED) and Quantum Chromodynamics (QCD).
Electroweak effects are covered in \cite{Fabiola}.

\subsection{Running Couplings}

\begin{figure}[b] 
\begin{center}
\includegraphics[height=3cm]{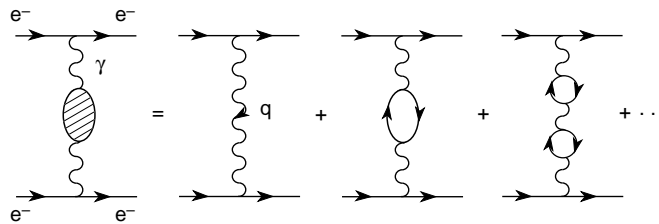}
\end{center}
\caption{Photon self-energy contribution to $e^-e^-$ scattering}
\label{fig:EEint}
\end{figure}

Let us consider the electromagnetic interaction between two electrons.
At lowest order, the scattering amplitude
$T(q^2)\sim \alpha/q^2$ with $\alpha = e^2/(4\pi)$.
The leading quantum corrections are generated by
the photon self-energy contribution:
$$ 
T(Q^2) \;\sim\; {\alpha\over Q^2} \, 
\left\{ 1 -\Pi(Q^2) + \Pi(Q^2)^2 + \cdots\right\}\; = \;
{\alpha\over Q^2} \, {1\over 1 + \Pi(Q^2)}
\;\sim\; {\alpha(Q^2)\over Q^2}\; .
$$ 
This defines an effective {\it running} coupling,
\begin{equation}\label{eq:alpha_run}
\alpha(Q^2) \; = \; {\alpha(Q_0^2)\over 1-{\beta_1\over 2\pi}\,
\alpha(Q_0^2)\,\ln{(Q^2/Q_0^2)}} \; ,
\end{equation}
%
\begin{figure}[tbh] 
\begin{center}
\includegraphics[height=6cm]{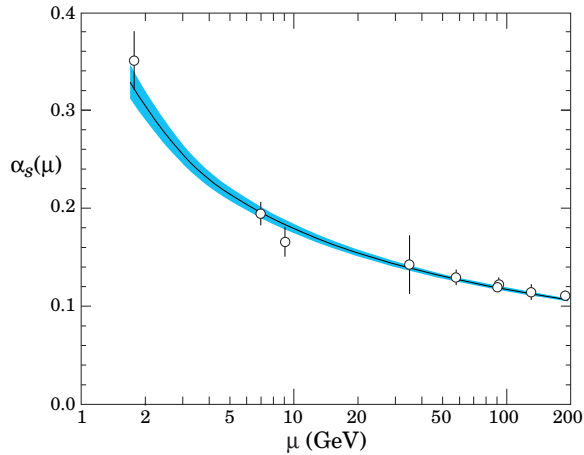}
\end{center}
\caption{Energy dependence \cite{PDG} of the strong coupling $\alpha_s$}
\label{fig:as_running}
\end{figure}
%
\hskip -0.1cm
where\ $Q^2\equiv -q^2$\ and\ $\alpha(m_e^2)=\alpha$.
The $e^+e^-$ loop induces a logarithmic correction with
$\beta_1= 2/3 >0$. Therefore, the effective QED running coupling increases 
with the energy scale:
$\alpha(Q^2)>\alpha(Q_0^2)$ if $Q^2>Q_0^2\,$,
i.e., the electromagnetic charge decreases at large distances. This can be
intuitively understood as the screening due to virtual $e^+e^-$ pairs
generated, through quantum effects, around the electron charge.
The physical QED vacuum behaves as a polarized dielectric medium.
The huge difference between the electron and $Z$ mass scales makes
this quantum correction relevant at LEP energies \cite{JE:01,BP:01}:
\begin{equation}\label{eq:alpha}
\alpha(m_e^2)^{-1}\; =\; 137.03599976\, (50)
\;\; > \;\;
\alpha(M_Z^2)^{-1}\; =\; 128.95\pm 0.05\; .
\end{equation}

The strong interaction between two quarks
can be analyzed in a similar way. 
Owing to the non-abelian character of the $SU(3)_C$ group,
QCD leads to cubic and quartic self-interactions among gluons. 
This results in a strong running coupling
$\alpha_s(Q^2)$ with the same $Q^2$ dependence (\ref{eq:alpha_run}),
but with a negative $\beta_1$ \cite{GW:73}:
\begin{equation}\label{eq:QCD_beta}
\beta_1 \; = \; {2\, N_f - 11\, N_C \over 6} \; < \; 0
\; .
\end{equation}
The contribution proportional to
the number of quark flavours $N_f$ is generated by the
$q$-$\bar q$ loop corrections to the gluon self-energy.
The gluonic self-interactions introduce the additional negative
term proportional to the number of quark colours $N_C$. 
Since $\beta_1 < 0$, $\alpha_s(Q^2)$
decreases at short distances.
Thus, QCD has the required property of \emph{asymptotic freedom}:
quarks behave as free particles when $Q^2\to\infty$.
The predicted running of $\alpha_s$, known to four loops \cite{beta},
agrees very well with the experimental determinations at
different energies. Normalizing all measurements at the $Z$ mass scale,
the present world average is \cite{PDG}:
\begin{equation}\label{eq:alpha_s}
\alpha_s(M_Z^2)\; =\; 0.118 \pm 0.002 \; .
\end{equation}
%

\begin{figure}[tbh] 
\begin{center}
\includegraphics[width=9.75cm]{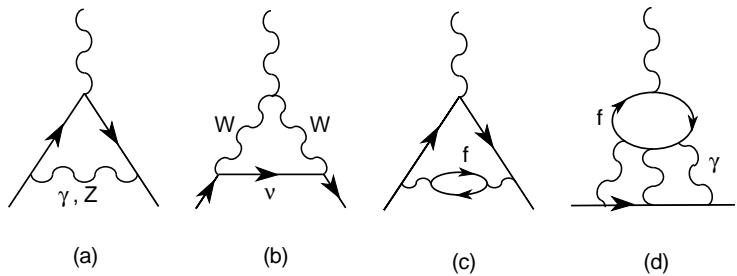}
\end{center}
\caption{Some Feynman diagrams contributing to $a_l^\gamma$}
\label{fig:AnMagMom}
\end{figure}
 
\subsection{Lepton Anomalous Magnetic Moments}

The most stringent QED test \cite{KI:96,CM:01}
comes from the high-precision
measurements of the $e$ \cite{PDG} and $\mu$ \cite{BNL:E821}
anomalous magnetic moments  $a_l^\gamma\equiv (g_l-2)/2$:
\begin{subeqnarray}
a_e^\gamma & = &\left\{\begin{array}{ccc}
(115 \, 965 \, 215.35\pm 2.40) \times 10^{-11} &\qquad & (\mathrm{Theory})
\\
(115 \, 965 \, 218.69\pm 0.41) \times 10^{-11} &\qquad & (\mathrm{Experiment})
\end{array}\right.\; , \label{eq:a_e}\\   
a_\mu^\gamma & = &\left\{ \begin{array}{ccc}
(1 \, 165 \, 917.9\pm 1.0)  \times 10^{-9} &\qquad & (\mathrm{Theory})
\\ 
(1 \, 165 \, 920.3\pm 1.5) \times 10^{-9} &\qquad & (\mathrm{Experiment})
\end{array}\right. \; . \label{eq:a_mu} 
\end{subeqnarray}
The impressive agreement between theory and experiment  
promotes QED 
to the level of the best theory ever build by the human mind 
to describe nature. 
 
To a measurable level, $a_e^\gamma$ arises entirely from virtual electrons and 
photons; these contributions are known \cite{KI:96} to $O(\alpha^4)$.  
The theoretical error is dominated by the uncertainty in the 
input value of the QED coupling $\alpha$.
Turning things around, $a_e^\gamma$ provides the most precise 
determination of the fine structure constant.
 
The anomalous magnetic moment of the muon is sensitive to virtual  
contributions from heavier states; compared to $a_e^\gamma$, they scale 
as $m_\mu^2/m_e^2$. 
The main theoretical uncertainty on $a_\mu^\gamma$ has a QCD origin. 
Since quarks have electric charge, virtual quark-antiquark pairs 
induce
{\it hadronic vacuum polarization} corrections to the photon propagator 
(Fig.~\ref{fig:AnMagMom}.c).
Owing to the non-perturbative character of QCD at low energies, 
the light-quark contribution cannot be reliably calculated at 
present; fortunately, this effect can be 
extracted from the measurement of the cross-section 
$\sigma(e^+e^-\to \mbox{\rm hadrons})$ 
and from the invariant-mass distribution of the final hadrons in 
$\tau$ decays \cite{JE:01}.
Additional QCD uncertainties stem from the smaller 
{\it light-by-light scattering} contributions (Fig.~\ref{fig:AnMagMom}.d);
a recent reevaluation of these corrections \cite{light_new} has detected 
a sign mistake in previous calculations \cite{light_old}, improving the
agreement with the experimental measurement \cite{BNL:E821}.

The Brookhaven E821 experiment \cite{BNL:E821} is expected to push
its sensitivity to at least $4\times 10^{-10}$,
and thereby observe the contributions from virtual $W^\pm$ and $Z$ bosons 
\cite{CM:01,a_mu_EW}.
This would require a better control of the QCD corrections.

\section{Lepton Universality}

In the SM all lepton doublets have identical couplings to the $W$ boson:
\begin{equation}\label{eq:W_ln}
{\cal L}\; =\; {g\over 2\sqrt{2}}\; W_\mu^\dagger\;\sum_l \; 
\bar\nu_l\,\gamma^\mu (1-\gamma_5)\, l \, +\, \mathrm{h.c.}
\qquad\qquad (l=e,\,\mu ,\,\tau )\; .
\end{equation}
%
\begin{table}[htb]
\caption{Experimental determinations of the ratios $g_l/g_{l'}$}
\begin{center}
\renewcommand{\arraystretch}{1.4}
\setlength\tabcolsep{5pt}
\begin{tabular}{llll}
\hline\noalign{\smallskip}
& $\Gamma_{\tau\to\nu_\tau\mu\,\bar\nu_\mu/\nu_\tau e\,\bar\nu_e}$ &
$\Gamma_{\pi\to\mu\,\bar\nu_\mu / e\,\bar\nu_e}$ &
$\Gamma_{W\to\mu\,\bar\nu_\mu / e\,\bar\nu_e}$
\\ \noalign{\smallskip}\hline\noalign{\smallskip}
$|g_\mu/g_e| =$ 
& $1.0006\pm 0.0021$ & $1.0017\pm 0.0015$ & $1.000\pm 0.011$
\\ \hline
\end{tabular}
\end{center}
\begin{center}
\renewcommand{\arraystretch}{1.4}
\setlength\tabcolsep{5pt}
\begin{tabular}{lllll}
\hline\noalign{\smallskip}
& $\Gamma_{\tau\to\nu_\tau e\,\bar\nu_e}/\Gamma_{\mu\to\nu_\mu e\,\bar\nu_e}$ &
$\Gamma_{\tau\to\nu_\tau\pi}/\Gamma_{\pi\to\mu\,\bar\nu_\mu}$ &
$\Gamma_{\tau\to\nu_\tau K}/\Gamma_{K\to\mu\,\bar\nu_\mu}$ &
$\Gamma_{W\to\tau\,\bar\nu_\tau/\mu\,\bar\nu_\mu}$
\\ \noalign{\smallskip}\hline\noalign{\smallskip}
$|g_\tau/g_\mu| =$ & $0.9995\pm 0.0023$ & $1.005\pm 0.007$ &
$0.977\pm 0.016$ & $1.026\pm 0.014$ 
\\ \hline
\end{tabular}
\end{center}
\begin{center}
\renewcommand{\arraystretch}{1.4}
\setlength\tabcolsep{5pt}
\begin{tabular}{lll}
\hline\noalign{\smallskip}
& $\Gamma_{\tau\to\nu_\tau \mu\,\bar\nu_\mu}/\Gamma_{\mu\to\nu_\mu e\,\bar\nu_e}$ &
$\Gamma_{W\to\tau\,\bar\nu_\tau/e\,\bar\nu_e}$
\\ \noalign{\smallskip}\hline\noalign{\smallskip}
$|g_\tau/g_e| =$ & $1.0001\pm 0.0023$ & $1.026\pm 0.014$ 
\\ \hline
\end{tabular}
\end{center}
\label{tab:univtm}
\end{table}
%
Comparing the measured decay widths of leptonic or semileptonic decays
which only differ by the lepton flavour, one can test experimentally
that the $W$ interaction is indeed the same, i.e. that \
$g_e = g_\mu = g_\tau \equiv g\, $.
As shown in Table~\ref{tab:univtm}, the present data
\cite{capri:97,PDG,LEPEWWG}
verify the universality of the leptonic charged-current couplings
to the 0.2\% level.

\begin{figure}[bt] 
\begin{center}
\includegraphics[height=7.5cm]{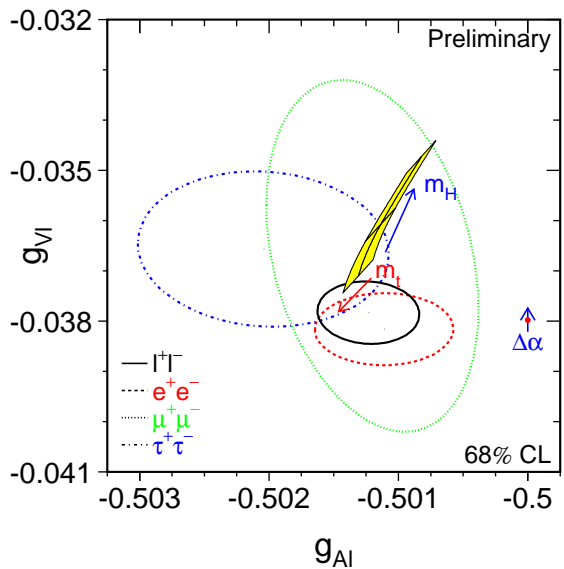}
\end{center}
\caption{Contours of 68\% probability in the $a_l$-$v_l$ plane from
LEP and SLD measurements \cite{LEPEWWG}. The solid contour assumes
lepton universality. The shaded region corresponds to the SM prediction
for $m_t =174.3\pm 5.1$~GeV and $m_H = 300\,{}^{+700}_{-186}$~GeV.
}
\label{fig:gal_gvl_contours}
\end{figure}

The interactions of the neutral $Z$ boson are diagonal in flavour.
Moreover, all fermions with equal electric charge have identical
axial-vector, $a_f = T_3^f=\pm 1/2$, and  
vector, $v_f = T_3^f\, (1-4\, |Q_f|\,\sin^2{\theta_W})$,
couplings to the $Z$.
This has been accurately tested at LEP and SLD through a
precise analysis of $e^+e^-\to\gamma,Z\to f\bar f$ data.
Figure~\ref{fig:gal_gvl_contours} shows the 68\% probability 
contours in the
$a_l$-$v_l$ plane, obtained from leptonic observables \cite{LEPEWWG}.
The universality of the leptonic $Z$ couplings is now verified to 
the $0.15\%$
level for $a_l$, while only a few per cent precision has been achieved 
for $v_l$ due to the smallness of the leptonic vector coupling.
The measured leptonic asymmetries provide
an accurate determination of the electroweak mixing angle \cite{LEPEWWG}:
\begin{equation}
\sin^2{\theta_W}\; = \; 0.23113\pm 0.00021\; .
\end{equation}

\section{Flavour Mixing}

In the SM, all mass scales are generated through the Higgs mechanism.
After the SSB, the Yukawa couplings to the Higgs scalar doublet give rise
to non-diagonal fermionic mass terms. The mass eigenstates are then
different from the weak eigenstates, which leads to flavour mixing
in the charged-current interaction:
\begin{equation}\label{eq:quark_mixing}
{\cal L}\; =\; {g\over 2\sqrt{2}}\; W_\mu^\dagger\;\sum_{ij} \; 
\bar u_i\,\gamma^\mu (1-\gamma_5)\,\mathbf{V}_{\! ij}\, d_j
\, +\,\mathrm{h.c.}\; .
\end{equation}
With non-zero neutrino masses, there are analogous mixing effects in the 
lepton sector, which are covered in \cite{Pilar}.

The Cabibbo-Kobayashi-Maskawa \cite{CA:63,KM:73} (CKM) 
matrix\ $\mathbf{V}$ is unitary and
couples any up-type quark with all down-type quarks. It is a priori unknown,
because the gauge symmetry does not fix the Yukawa couplings.
The matrix element $\mathbf{V}_{\! ij}$ can be obtained experimentally from
semileptonic weak processes associated with the quark transition
$d_j\to u_i l^-\bar\nu_l$. The present determinations are
summarized in Table~\ref{tab:V_CKM}. The uncertainties are
dominated by theoretical errors, related to the strong interaction
which binds quarks into hadrons.

\begin{table}[bth]
\caption{Direct $\mathbf{V}_{\! ij}$ determinations.}
\begin{center}
\renewcommand{\arraystretch}{1.4}
\setlength\tabcolsep{10pt}
\begin{tabular}{lll}
\hline\noalign{\smallskip}
CKM entry & Value & Source
\\ \noalign{\smallskip}\hline\noalign{\smallskip}
$|\mathbf{V}_{\! ud}|$ & $0.9740\pm 0.0010$ & Nuclear $\beta$ decay 
\cite{PDG}
\\
& $0.9733\pm 0.0015$ &
$n\to p\, e^-\bar\nu_e$ \cite{PDG}
\\ \hline 
$|\mathbf{V}_{\! us}|$ & $0.2196\pm 0.0023$ & $K_{e3}$ \cite{PDG}
\\
& $0.2176\pm 0.0026$ & Hyperon decays \cite{PDG}
\\ \hline
$|\mathbf{V}_{\! cd}|$ & $0.224\pm 0.016$ & $\nu\, d \to c\, X$ 
\cite{PDG}
\\ \hline
$|\mathbf{V}_{\! cs}|$ & $1.04\pm 0.16$ & $D \to \bar K\, e^+\nu_e$  
\cite{PDG}
\\ \hline
$|\mathbf{V}_{\! cb}|$ & $0.0421\pm 0.0022$ & $B\to D^* l\bar\nu_l$ 
\cite{AB:02}
\\
& $0.0404\pm 0.0011$ & $b\to c\, l\,\bar\nu_l$
\cite{AB:02}
\\ \hline
$|\mathbf{V}_{\! ub}|$ & $0.0033\pm 0.0006$ &
 $B\to\rho\, l\,\bar\nu_l$ \cite{CLEO:00}
\\
& $0.0041\pm 0.0006$ & $b\to u\, l\,\bar\nu_l$ \cite{HFG,CLEO:02}
\\ \hline
\raisebox{-.75ex}{
$|\mathbf{V}_{\! tb}|\, / \sqrt{\sum_q |\mathbf{V}_{\! tq}|^2}$}
& \raisebox{-.75ex}{$0.97\, {}^{+0.16}_{-0.12}$} & 
\raisebox{-.75ex}{$t\to b\, W / q\, W$ \cite{CDF:01}}
\\[7pt]  \hline
\end{tabular}
\end{center}
\label{tab:V_CKM}
\end{table}

The most precisely known CKM matrix element is $\mathbf{V}_{\! ud}$.
The weighted average of the two determinations in Table~\ref{tab:V_CKM} gives
$\mathbf{V}_{\! ud}= 0.9738\pm 0.0008$~. Taking for $\mathbf{V}_{\! us}$
the more reliable $K_{e3}$ determination, one obtains
\begin{equation}\label{eq:unit}
|\mathbf{V}_{\! ud}|^2 \, +\, |\mathbf{V}_{\! us}|^2 \, 
+\, |\mathbf{V}_{\! ub}|^2 \; =\; 0.9965\pm 0.0019\; .
\end{equation}
The unitarity of $\mathbf{V}_{\! ij}$ 
appears to be slightly violated by $1.8\sigma$.
At this level of precision, 
a small underestimate of some uncertainties seems plausible. 
A less accurate unitarity test is
provided by the hadronic width of the $W$ boson \cite{LEPEWWG}:
\begin{equation}\label{eq:W_had}
\sum_{j=d,s,b}\; \left( |\mathbf{V}_{\! uj}|^2 \, 
+\, |\mathbf{V}_{\! cj}|^2 \right) \; = \; 2.039\pm 0.025\; .
\end{equation}

The CKM matrix shows a hierarchical pattern, with its
diagonal elements being very close to one, the ones connecting the
two first generations having a size
$\lambda\equiv |\mathbf{V}_{\! us}|$,
the mixing between the second and third families being of order
$\lambda^2$, and the mixing between the first and third quark flavours
having a much smaller size of about $\lambda^3$.
It is convenient to use the
parameterization \cite{WO:83}:
\begin{equation}\label{eq:wolfenstein}
\mathbf{V}\; =\;\left[
\setlength\arraycolsep{5pt}
\begin{array}{ccc}
1- {\lambda^2 / 2} & \lambda & A\lambda^3(\varrho  - i\eta) \\
-\lambda & 1 -{\lambda^ 2 / 2} & A\lambda^ 2 \\
A\lambda^ 3(1-\varrho -i\eta) & -A\lambda^ 2 & 1 \end{array}
\right] \; + \; O(\lambda^4)\; .
\end{equation}
Imposing the unitarity constraint, the CKM determinations in 
Table~\ref{tab:V_CKM} imply
\begin{equation}\label{eq:circle}
\lambda = 0.223\pm 0.003 \; , \qquad
A =  0.83\pm 0.04 \; ,\qquad
\sqrt{\varrho^2+\eta^2}  =  0.40\pm 0.07 \; .
\end{equation}

\subsection{$B^0$-$\bar B^0$ Mixing}

\begin{figure}[htb] 
\begin{center}
\includegraphics[height=2.5cm]{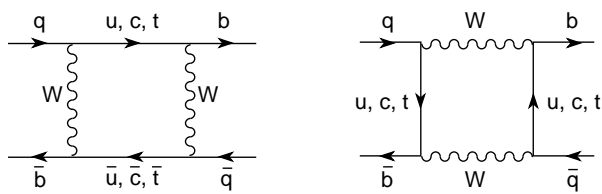}
\end{center}
\caption{$B^0$-$\bar B^0$ mixing diagrams}
\label{fig:boxdia}
\end{figure}

Additional information on the CKM parameters is obtained from
flavour-changing neutral-current transitions, occurring at the 1-loop
level. An important example is provided by 
the mixing between the $B^0$ meson and its antiparticle:
\begin{equation}\label{eq:mixing}
\langle\bar B_d^0 | {\cal H}_{\Delta B=2}|B_0\rangle\,\sim\,
\left\{\sum_{ij=u,c,t}\, \mathbf{V}_{\! id}^{\phantom{*}}
\mathbf{V}_{\! ib}^*\mathbf{V}_{\! jd}^*\mathbf{V}_{\! jb}^{\phantom{*}}\;
S(r_i,r_j) \right\} \;
\left( {2\over 3} M_B^2 \xi_B^2\right)
\; ,
\end{equation}
with $S(r_i,r_j)$ a loop function of $r_i\equiv m_i^2/M_W^2$. 
Owing to the unitarity of the CKM matrix, the mixing amplitude vanishes
for equal (up-type) quark masses. 
Thus the effect
is proportional to the mass splittings between the $u$, $c$ and $t$ quarks.
Since all CKM factors have a similar size,
$\mathbf{V}_{\! ud}^{\phantom{*}}\mathbf{V}_{\! ub}^*\sim
\mathbf{V}_{\! cd}^{\phantom{*}}\mathbf{V}_{\! cb}^*\sim
\mathbf{V}_{\! td}^{\phantom{*}}\mathbf{V}_{\! tb}^*\sim A\lambda^3$,
the top contribution dominates completely.
This transition can then be used to perform an indirect determination
of $|\mathbf{V}_{\! td}|$.
The main uncertainty stems from the hadronic
matrix element of the four-quark operator
generated by the box diagrams in Fig.~\ref{fig:boxdia}, 
which is characterized through the non-perturbative parameter
$\xi_B\equiv \sqrt{2\hat B_B}\,f_B = 230 \pm 45$~MeV
\cite{latt,SR}.
The measured mixing between the $B_d^0$ and $\bar B_d^0$ mesons,
$\Delta M_{B^0_d} = 0.496\pm 0.007$ ps$^{-1}$  \cite{HFG},
implies:
\begin{equation}\label{eq:V_td}
 |\mathbf{V}_{\! td}|\, = \, 0.0077\pm 0.0011\; ,
\qquad
\sqrt{(1-\varrho)^2+\eta^2} \, \approx \,
\left|{\mathbf{V}_{\! td}\over \lambda\mathbf{V}_{\! cb}}\right|
\, = \, 0.84\pm 0.12\; .
\end{equation}

A similar analysis can be applied to the $B^0_s$-$\bar B^0_s$ mixing.
The non-perturbative uncertainties are reduced to the
level of $SU(3)$ breaking through the ratio
\begin{equation}\label{x_ratio}
{\Delta M_{B^0_s}\over \Delta M_{B^0_d}} \;\approx\;
{M_{B^0_s}\, \xi^2_{B^0_s}\over M_{B^0_d}\, \xi^2_{B^0_d}}\;
\left|{\mathbf{V}_{\! ts}\over \mathbf{V}_{\! td}}\right|^2
\;\equiv\; \Omega^2 \;
\left|{\mathbf{V}_{\! ts}\over \mathbf{V}_{\! td}}\right|^2 \, .
\end{equation}
Taking $\Omega\approx 1.15\pm 0.08$ \cite{latt,SR},
the experimental bound $\Delta M_{B^0_s} > 14.9$~ps$^{-1}$ 
(95\%\,\mbox{\rm CL}) \cite{HFG}
implies
\begin{equation}\label{eq:Vts_Vtd}
\left|{\mathbf{V}_{\! ts}\over \mathbf{V}_{\! td}}\right|
\;\approx\; {1\over \lambda  \sqrt{(1-\varrho)^2+\eta^2}}\; >\; 4.2 \; .
\end{equation}

\section{CP Violation}

With $N_G$ fermion generations, the matrix
$\mathbf{V}$ is characterized by
$N_G(N_G-1)/2$ moduli and $(N_G-1)(N_G-2)/2$ phases.
In the simpler case of two fermion families $\mathbf{V}$
is determined by a single parameter, the so-called
Cabibbo angle~\cite{CA:63}, while for $N_G=3$ the CKM matrix 
is described by 3 angles and 1 phase~\cite{KM:73}.
This is the only complex phase in the SM Lagrangian;
thus, it is a unique source for violations of the CP symmetry.
It was for this reason that the third generation
was assumed to exist \cite{KM:73},
before the discovery of the $\tau$ and the $b$.

\subsection{Kaon Physics}

For many years, the only experimental evidence of CP-violation
phenomena came from the kaon system. The ratios,
\begin{equation}\label{eq:eta}
{A(K_L\to\pi^+\pi^-)\over A(K_S\to\pi^+\pi^-)}
 \,\approx\, \varepsilon_K^{\phantom{'}} + \varepsilon_K' 
\; , \qquad
{A(K_L\to\pi^0\pi^0)\over A(K_S\to\pi^0\pi^0)}
 \,\approx\, \varepsilon_K^{\phantom{'}} - 2\varepsilon_K' \; ,
\end{equation}
involve final $2\pi$  states which are even under CP. Therefore, they
measure a CP-violating amplitude which can originate
either from a small CP-even admixture in the initial $K_L$ state
(indirect CP violation), parameterized by $\varepsilon_K$, or from
direct CP violation in the decay amplitude. This latter effect,
parameterized by $\varepsilon_K'$, requires the interference between
the two $K\to 2\pi$ isospin ($I=0,2$)  amplitudes, with
different weak and strong phases.

The parameter $\varepsilon_K$ is well determined \cite{PDG}:
\begin{equation}\label{eq:eps_K}
\varepsilon_K  =  (2.271\pm 0.017)\times 10^{-3}\;\;
     \mathrm{e}^{i\phi(\varepsilon_K)} \; ,
\qquad
\phi(\varepsilon_K) = 43.5^\circ\pm 0.5^\circ\;  .
\end{equation}
$\varepsilon_K$ has been also measured \cite{PDG,KTeVs} through the
CP asymmetry between the two
$K_L\to\pi^\mp l^\pm\stackrel{{}_{(-)}}{\nu_l}$ 
decay widths, which implies 
Re$\, (\varepsilon_K) = (1.654\pm0.032)\times 10^{-3}$, in good agreement
with (\ref{eq:eps_K}).

The value of $\varepsilon_K'$ has been established very recently.
The present experimental world average \cite{NA48,NA31,KTeV,E731},
\begin{equation}\label{eq:eps'}
\mathrm{Re}\, (\varepsilon'_K/\varepsilon^{\phantom{'}}_K) 
 = (1.72\pm 0.18)\times 10^{-3} \; ,
\end{equation}
provides clear evidence for the existence of direct CP violation.

The CKM mechanism generates CP-violation effects both in the
$\Delta S=2\; $ $K^0$-$\bar K^0$ transition (box diagrams) and in the
$\Delta S=1$ decay amplitudes (penguin diagrams).
The theoretical analysis of $K^0$-$\bar K^0$ mixing is
quite similar to the one applied to the $B$ system. This time, however,
the charm loop contributions are non-negligible. The main
uncertainty stems from the calculation of the hadronic matrix
element of the four-quark $\Delta S=2$ operator, which is usually
parameterized through the non-perturbative parameter $\hat B_K$.

\begin{figure}[tbh] 
\begin{center}
\includegraphics[height=5.25cm]{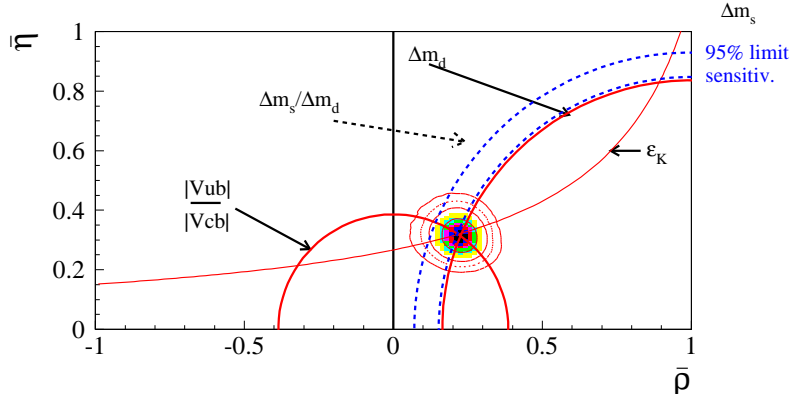}
\end{center}
\caption{Constraints on the $(\varrho,\eta)$ vertex of the
unitarity triangle \protect\cite{CKM:02}}
\label{fig:unitarity_constraints}
\end{figure}

The experimental value of $\varepsilon_K$ specifies a hyperbola in the
$(\varrho,\eta)$ plane.
This is shown in Fig.~\ref{fig:unitarity_constraints} \cite{CKM:02},
together with the constraints obtained from
$|\mathbf{V}_{\! ub}/\mathbf{V}_{\! cb}|$,
$B_d^0$-$\bar B_d^0$ mixing and the experimental bound on
$\Delta M_{B_s^0}$.
This figure assumes \cite{CKM:02} 
$\hat B_K = 0.87\pm 0.06\pm 0.13$, $\xi_B = 230\pm 25\pm 20$ MeV
and $\Omega = 1.15\pm 0.04\pm 0.05$.

The theoretical estimate of $\varepsilon'_K/\varepsilon^{\phantom{\prime}}_K$
is more involved \cite{Buras}, 
because several four-quark operators need to be considered
in the analysis. Moreover, the strong rescattering of the final pions
generates an important enhancement through infrared logarithms \cite{PPS:01}.
Taking into account all large logarithmic corrections at short
and long distances, the SM prediction for
$\varepsilon'/\varepsilon$ is found to be \cite{PPS:01}:
\begin{equation}\label{eq:SMpred}
\mathrm{Re}\left(\varepsilon'/\varepsilon\right) \; =\;  
\left(1.7\pm 0.2\, {}_{-0.5}^{+0.8} \pm 0.5\right) 
\times 10^{-3}\; ,
\end{equation}
in excellent agreement with the measured experimental
value (\ref{eq:eps'}).

\subsection{$B$ Physics}

The unitarity tests in (\ref{eq:unit}) and (\ref{eq:W_had})
involve only the moduli of the CKM parameters, while 
CP violation has to do with their phases.
The most interesting off-diagonal unitarity condition is
\begin{equation}\label{eq:triangle}
\mathbf{V}^\ast_{\! ub}\mathbf{V}^{\phantom{*}}_{\! ud}  + 
\mathbf{V}^\ast_{\! cb}\mathbf{V}^{\phantom{*}}_{\! cd}  + 
\mathbf{V}^\ast_{\! tb}\mathbf{V}^{\phantom{*}}_{\! td}  =  0 \, ,
\end{equation}
which has three terms of a similar size.
This relation can be visualized by a triangle in the complex
plane, which is usually
scaled by dividing its sides by 
$\mathbf{V}^\ast_{\! cb}\mathbf{V}^{\phantom{*}}_{\! cd}$.
This aligns one
side of the triangle along the real axis and makes its length equal to
1; the coordinates of the 3 vertices are then
$(0,0)$, $(1,0)$ and $(\varrho,\eta)$.
In the absence of CP violation ($\eta = 0$), this unitarity triangle 
would degenerate into a segment along the real axis.

The length of the sides and the angles 
($\alpha$, $\beta$, $\gamma$) of the unitarity triangle
can be directly measured.
In fact, we have already determined its sides from 
$\Gamma(b\to u)/\Gamma(b\to c)$ and 
$B^0_d$-$\bar B^0_d$ mixing, and the position of the $(\varrho,\eta)$
vertex has been further pinned down in Fig.~\ref{fig:unitarity_constraints}
with 
$\varepsilon_K$.
This gives \cite{CKM:02}:
\begin{equation}\label{eq:CKM_fit}
\varrho = 0.224\pm 0.038\; , \qquad
\eta = 0.317\pm 0.040\; , \qquad
\sin{2\beta} = 0.698\pm 0.066\; ,
\end{equation}
where
$\beta\equiv -\arg(\mathbf{V}^{\phantom{*}}_{\! cd}\mathbf{V}^*_{\! cb}/
\mathbf{V}^{\phantom{*}}_{\! td}\mathbf{V}^*_{\! tb})$.

$B^0$ decays into CP self-conjugate final states
provide independent ways to determine the angles \cite{CPself}.
The $B^0$ (or $\bar B^0$) can decay directly to the given final state $f$,
or do it after the meson has been changed to its antiparticle via the
mixing process. CP-violating effects can then result from the interference
of these two contributions. 
The time-dependent CP-violating rate asymmetries 
contain direct information on the CKM parameters.
The gold-plated decay mode is $B^0_d\to J/\psi K_S$, which 
gives a clean measurement of $\beta$ \cite{SinBeta} without
strong-interaction uncertainties,
in good agreement with (\ref{eq:CKM_fit}): 
\begin{equation}\label{eq:beta}
\sin{2\beta} = 0.80\pm 0.10\; .
\end{equation}

Additional tests of the CKM matrix are underway.
The $B$ factories should accomplish an approximate determination of
$\alpha\equiv -\arg(
   \mathbf{V}^{\phantom{*}}_{\! td}\mathbf{V}^*_{\! tb}/ 
   \mathbf{V}^{\phantom{*}}_{\! ud}\mathbf{V}^*_{\! ub})$,
from $ B^0_d\to\pi^+\pi^-$, and many other interesting
studies with $B$ decays.
Complementary and very valuable information could be also
obtained from the kaon decay modes
$K^\pm\to\pi^\pm\nu\bar\nu$, $K_L\to\pi^0\nu\bar\nu$ and
$K_L\to\pi^0e^+e^-$ \cite{jaca:94,Buras}.

\section{Summary}

The SM provides a beautiful theoretical framework which is able to
accommodate all our present knowledge on electroweak and strong
interactions. It is able to explain any single experimental fact  
and, in some cases, it has successfully passed very precise
tests at the 0.1\% to 1\% level \cite{Fabiola}. 
In spite of this impressive phenomenological success, the SM leaves 
too many unanswered questions to be considered as a complete 
description of the fundamental forces. We do not understand yet 
why fermions are replicated in three (and only three)
nearly identical copies? Why the pattern of masses and mixings
is what it is?  Are the masses the only difference among the three
families? What is the origin of the SM flavour structure?
Which dynamics is responsible for the observed CP violation?

The fermionic flavour is the main source of arbitrary free 
parameters in the SM. The problem of fermion-mass
generation is deeply related with the mechanism responsible for the 
electroweak SSB.
Thus, the origin of these parameters lies in the most obscure part of
the SM Lagrangian: the scalar sector. 
Clearly, the dynamics of flavour appears to be ``terra incognita''
which deserves a careful investigation.

The SM incorporates a mechanism to generate CP violation, through the
single phase naturally occurring in the CKM matrix.
Although the present laboratory experiments are well described,
this mechanism is unable to explain the
matter-antimatter asymmetry of our universe.
A fundamental explanation of the origin of CP-violating phenomena is
lacking.

Many interesting experimental signals are expected to be seen in the
near future. Large surprises may well be
discovered, probably giving the first hints of new physics and
offering clues to the problems of mass generation, fermion
mixing and family replication.

\section*{Acknowledgements}

This work has been supported by MCYT, Spain 
(Grant FPA-2001-3031) and by the EU TMR network
EURODAPHNE (Contract ERBFMX-CT98-0169).

%

\end{document}